\documentstyle[12pt,aaspp4]{article}

\def\cite{{\it cite}}

\def\gtsim{\ {\raise-0.5ex\hbox{$\buildrel>\over\sim$}}\ }
\def\ltsim{\ {\raise-0.5ex\hbox{$\buildrel<\over\sim$}}\ }
\slugcomment{Submitted to The Astrophysical Journal}

\begin{document}

\title{The RR Lyrae Population of the Galactic Bulge from the MACHO Database:
Mean Colors and Magnitudes}

\author {C.  Alcock$^{1,2}$, R.A.  Allsman$^{1,4}$, D. R. Alves$^{1,9}$,
T.S. Axelrod$^{1,4}$, A.C. Becker$^{2,3}$, A. Basu$^{1}$, L. Baskett$^{1}$, 
D.P. Bennett$^{1,2}$, K.H. Cook$^{1,2}$, 
K.C. Freeman$^4$, K. Griest$^{2,5}$, J.A. Guern$^5$, M.J. Lehner$^5$, 
S.L. Marshall$^{1}$, D. Minniti$^{1}$,
B.A. Peterson$^4$, M.R. Pratt$^{2,3}$, P.J. Quinn$^6$, A.W. Rodgers$^4$, C.W.
Stubbs$^{2,3,6}$, W. Sutherland$^7$, T. Vandehei$^{5}$, and D.L. Welch$^{8}$}
\affil { (The MACHO Collaboration) }

\altaffiltext{1}{Lawrence Livermore National Laboratory, Livermore, CA 94550\\
E-mail:  alcock, abasu, lbaskett, robynallsman, tsa, bennett, kcook, dminniti, stuart@llnl.gov}

\altaffiltext{2}{Center for Particle Astrophysics, University of California,
Berkeley, CA 94720}

\altaffiltext{3}{Department of Astronomy, University of Washington,
Seattle, WA 98195 \\
E-mail: stubbs, becker@astro.washington.edu}

\altaffiltext{4}{Mt Stromlo \& Siding Spring Observatories, Australian~National~Univ,~Weston~ACT~2611,~Australia \\ E-mail: kcf, peterson, 
pjq, alex@merlin.anu.edu.au}

\altaffiltext{5}{Department of Physics, University of California,
San Diego, CA 92093 \\
E-mail: griest, jguern, matt, vandehei@astrophys.ucsd.edu}

\altaffiltext{6}{European Southern Observatory, D-85748 Garching bei M\"unchen, Garmany\\
E-mail: pjq@eso.org}

\altaffiltext{7}{Department of Physics, University of Oxford,
Oxford OX1 3RH, U.K.\\
E-mail: wjs@oxds02.astro.ox.ac.uk}

\altaffiltext{8}{Department of Physics and Astronomy, McMaster University,
Hamilton, ON L8S 4M1, Canada\\
E-mail: welch@physics.mcmaster.ca}

\altaffiltext{9}{Department of Physics, University of California, Davis, CA 95616}

\begin{abstract}
Mean colors and magnitudes of RR Lyrae stars 
in 24 fields towards the Galactic bulge from the MACHO database are presented.  
Accurate mean reddenings are computed for these fields on the basis of
the mean colors. 
The distribution along the line of sight of the RR Lyrae population is examined
on the basis of the mean magnitudes, and it is shown that
the bulk of the RR Lyrae population is not barred. Only the RR Lyrae
in the inner fields closer to the Galactic center ($l<4^{\circ}, ~b>-4^{\circ}$)
show evidence for a bar. The red giant clump stars in the
MACHO fields, however, clearly show a barred distribution, confirming
the results of previous studies. Given the different spatial distribution,
the RR Lyrae and the clump giants trace two different populations.
The RR Lyrae would represent the inner extension of the Galactic halo
in these fields.
\end{abstract}

\section{Introduction}

There is a bar in the inner Milky Way, seen from the integrated IR light
(Blitz \& Spergel 1991, Dwek et al. 1995), 
from tracers of metal-rich populations such as Miras
(Whitelock 1994), AGB stars (Weinberg 1992), and RGB clump stars (Stanek et al. 1994), 
and from the kinematics of gas (Binney et al. 1991), and stars (Zhao et al. 1994).
This inner bar has the closest side at positive longitudes, and it extends 
as far as $l \approx 15^{\circ}$ (Dwek et al. 1995).
Several reviews have been written on the subject, that cover the
now extensive evidence (see the proceedings edited by Blitz \& Teuben 1996,
Buta et al. 1996, and Minniti \& Rix 1996).

Binney et al. (1996) have recently modeled
the photometric structure of the COBE/DIRBE maps. 
They find that the bar is almost pointing toward us ($\phi = 20^{\circ}$,
with axes ratios $1.0:0.6:0.4$, and semimajor axis size of 2 kpc.
This infrared emission is mostly due to stars.
There must be a clear magnitude difference between the stars located in
the near and far sides of the
bar, about $0.5\pm 0.2$ mag every 20 degrees, depending on the model
adopted.  The closer the alignment of the bar with the line connecting
the Sun and the Galactic center, the higher the expected magnitude difference. 

Clump giants have been used to study the spatial distribution of the Milky Way
bulge, and also show a barred distribution,
with the closer side of the bar located at positive longitudes.
Stanek et al. (1996) find a difference of 0.4 mag between the mean
mag of clump giants in their fields at $l=-5$ and $+5$ degrees.

The MACHO project has identified  large numbers of periodic 
variable stars in the bulge
fields (Cook et al. 1995). There are variable stars of 
different types, some of which
are good distance indicators (Miras, RR Lyrae, W UMas, $\delta$~Scu), with
well established period-luminosity relations. They
can be used as probes to study the density of  
different Galactic components (disk, bulge and halo)
along the line of sight towards the bulge.   
This is an important part of the MACHO project, because
the presence or absence of an inner Galactic
bar can significantly affect the optical 
depth to microlensing (e.g. Paczy\'nski et al. 1994).
Understanding the structure of the inner Galaxy and the interplay of the
different components is crucial in order to determine the masses of the 
lenses for the observed microlensing events. 

The RR Lyrae stars are by far the best distance 
indicators (e.g. Nemec et al. 1994), and are the focus of the present study. 
Because RR Lyrae stars are good distance indicators,
they can be used not only to measure the distance to the Galactic center
$R_0$, but also to
study the structure of the inner Milky Way.
In this paper we study the colors and magnitudes of about 1800 RR Lyrae stars 
found in the MACHO bulge fields.
Our main goal is to
test the presence of a barred distribution,
taking into account that the RR Lyrae stars trace the metal-poor stellar populations.  

\section{The Data}

The system and data collection of the MACHO experiment is described by 
Alcock et al. (1996).
The MACHO database contains so far 38718 periodic variable stars
in the Milky Way bulge. A description of the 
variable star database has been published elsewhere (Cook et al. 1995). 
The bulge RR Lyrae stars were 
selected using the period-amplitude and amplitude-amplitude diagrams,
as we have done with the LMC RR Lyrae (Alcock et al. 1995).
Here we analyze the data from 1993, covering over 100 days.  The selection 
of RR Lyrae in this database is complicated by aliases. Some of the stars with
periods $P = 1/n ~days$ maybe badly phased because the MACHO 
observations are done roughly every 24 hours. However, most 
aliases are discarded by demanding that $P_V=P_R$, which stresses the
importance of having simultaneous coverage in two passbands. 
The selection of RR Lyrae is also complicated by the presence of other 
variable stars with overlapping periods: eclipsing binaries, $\delta$ Scu stars,
etc.  Fortunately again, we have good quality light curves in two passbands, and
the amplitude ratios can be used to discriminate the pulsating variable
stars from the eclipsing variables. 

The final RR Lyrae star sample is then selected from the Bailey (period-amplitude)
diagram, including stars with $0.2^d < P < 1.1^d$, and $0.1 < A_V < 2$,
using the fact that $A_R < 0.8 A_V$.
The best quality light curves were fitted
by a Fourier series of sine functions (see Smith 1995). 
The resulting Fourier coefficients 
(e.g. the $R_{31}$ $vs.$ $\phi_{31}$ or the $\phi_{21}$ $vs.$ $\phi_{31}$ planes)
were used to discriminate RR Lyrae stars pulsating in the fundamental mode
(RRab), in the first-overtone (RRc), and in the second-overtone (RRe),
from eclipsing binary stars. This method works well, but it requires  good
sampling of the light curve, because the Fourier coefficients are very sensitive
to a small number of points or photometric errors. 
Finally, all light curves were visually inspected.
This is time consuming for such a
large sample, but it is a reliable method.

This final sample, though representative of the whole RR Lyrae population in the 
bulge, is not complete.
Independent estimates of the completeness of our RR Lyrae sample are obtained
by comparison with other surveys (Blanco 1984, Udalski et al. 1993).
Using our internal redundancy, we estimate that
our RR Lyrae type ab sample is 93\% complete within the period cuts selected.
However, this incompleteness does not influence the results of this paper,
because the RR Lyrae stars that we miss in the overlap regions
are not biased towards the fainter ones.

The photometric calibration of such a large database using non-standard
filters is challenging. We have used the latest calibrations 
(Alves et al. 1996), and have made a series of $external$ comparisons with other
photometry (Walker \& Mack 1986, Cook 1987, Udalski et al. 1993).
In particular, the zero point is fixed by the LMC calibration, and agrees
with these other previous photometry.

The Macho\_V (MV) and Macho\_R (MR) magnitudes are transformed to standard 
Cousins V and R magnitudes using the following equations:
$$V = 23.38+1.0026\times MV - 0.15\times (MV - MR)$$
$$R = 23.20+1.0044\times MR + 0.18\times (MV - MR).$$

Figure 1 shows the color-magnitude diagram for all the RR Lyrae stars in
our sample (V $vs.$ V--R). The effect of absorption
is clear in this diagram, as stars string along the direction predicted
by the reddening vector.

The majority of the RR Lyrae stars discovered here are located in the MACHO 
fields which are closer to the Galactic center, including Baade's window. 
The distribution on the sky of these RRab is shown in Figure 1 of 
Alcock et al. (1997a, Paper I).
The MACHO fields are located in the northern--most extension of the
field studied by Alard (1996a), who found an increasing number of
RR Lyrae stars towards the Galactic center. This trend is also seen in our fields.
Table 1 lists the position of our fields in equatorial and Galactic
coordinates, as well as an estimate of the total number of stars
photometered. Each field is $42$ arcmin on a side ($1/8$ of each field was
not photometered in one color due to a dead amplifier). The fields listed in
Table 1 are ordered in terms of the total expected optical depth to
microlensing.

Most of the RR Lyrae stars in the final sample belong to the Galactic bulge. 
Their magnitudes peak at V = 15, which places them at about 8 kpc.
A few, however, are more than twice as far away ($\sim$50 of these 
stars).  These distant RRab stars belong to the Sgr galaxy, and
have been studied by Alcock et al. (1997a, Paper I).
These stars are not considered here.

There are also RR Lyrae type e,
with mean periods $P = 0.28^d$, and amplitudes $A_V = 0.2-0.4$.
Similar stars were identified by Alcock et al. (1995) in the LMC.
In this analysis we include them in the RRc group,
since their mean magnitudes and colors are not significantly 
different.

\section{Reddening}

One of the first problems that we  encounter when analyzing the
MACHO variable star database is reddening.  The bulge fields have high
non-uniform absorption. The determination of mean reddenings in these
fields is important for the study of different kinds of variable
stars in the database. For example, W~UMa stars can be used as distance
indicators to trace the structure of the inner disk (see Rucinski 1995).
However, their unreddened magnitudes and colors must be known. Also,
knowledge of the reddenings will allow us to select the fields that
are windows, i.e. where we can see all the way through the bulge
given our magnitude limits.

We can take advantage of the fact that RR Lyrae lie in the instability strip, and that
the color range of the instability strip is sufficiently narrow, 
that we may measure the mean reddening in each field. We chose the 
amplitude-color diagram in order to illustrate the effect of reddening, 
since there is a small dependence of
amplitude with color (e.g. Bono \& Stellingwerf
1994).  Figure 2 shows the amplitude-color diagram for 
RRab stars in four selected fields: F125, which has low and uniform absorption,
F119 (Baade's window), which represents an intermediate case in terms of
amount and homogeneity of the absorption, and two extreme cases of
very obscured fields with significant differential reddening,
F108 and F101. In particular, note that two sequences appear in F101, a feature
that is common in other fields, due to partial absorption behind
a foreground cloud.

Table 2 lists the mean colors and magnitudes of bulge RRab and RRc.
The numbers of stars averaged in each case (between 30 and 100 per field) 
are also indicated in this Table.  Figure 3 shows the mean $V-R$ colors of the
RRab and RRc in each of the MACHO bulge fields as function of Galactic latitude.
Although there is a strong dependence of mean $V-R$ on Galactic latitude, as
expected (e.g. Blanco 1992), there is not much dispersion around this trend.
Figure 4 shows the mean $V-R$ colors of both the
RRab and RRc in each of the MACHO bulge fields as function of Galactic 
longitude. There is no systematic dependence of the color on the longitude,
in agreement with the conclusions of Wozniak \& Stanek (1996) that there
are no large variations in the reddening law across the bulge.

In order to determine $E_{B-V}$ from the mean $V-R$, we use Baade's window as zero point.
The mean absolute reddening in Baade's window is $E_{B-V} = 0.50 \pm 0.03$ (van den
Bergh 1971, Blanco \& Blanco 1985, Terndrup et al. 1995, Stanek 1996).
Using the standard extinction law of Rieke \& Lebofski (1985), 
$E_{V-R} = 0.78\times E_{B-V} = 0.39$. The observed mean color of RRab in Baade's window is
$V-R = 0.60$. This gives a mean dereddened color $V-R = 0.21$ for Baade's window RRab
($V-R = 0.10$ for RRc),
consistent with the predictions of recent models and observations
of RR Lyrae stars in Galactic globular clusters (e.g. Silbermann \& Smith 1995,
Reid 1996, Walker \& Nemec 1996).
Therefore, the mean reddenings in the different fields are obtained from RRab by
$E_{V-R} = (V-R)_{obs} - 0.20$ (for RRc $E_{V-R} = (V-R)_{obs} - 0.08$).  
The reddenings derived from RRab and RRc are very similar in all cases.
These mean reddenings are listed in column
12 of Table 2, and are the result of averaging the results from RRab and RRc. 
The fields F111, F116, and F125 are the ones with the lowest
and more uniform reddening. 
We note that Gould et al. (1997) and Alcock et al. (1997b) have recently 
derived a smaller zero point extinction value for Baade's window. Adopting 
this new correction $\Delta E_{V-K} = 0.1$ would decrease accordingly all 
the absolute reddening values adopted above. 

The reddening is patchy in the MACHO fields towards the bulge, 
ranging from E(B-V) = 0.2 in the clear regions of the 
outer fields, to E(B-V) $>$ 1 in the most obscured regions. 
In fact, the reddest RRab stars in the sample have $V-R \approx 1.0$,
which implies a large extinction, $A_V \approx 3.2$ mag.

\section{The Structure of the Inner Bulge: Evidence of a Bar}

\subsection{The RRab as tracers}

In this Section we will study the structure of the bulge using the mean
magnitudes of the RR Lyrae in the different fields in order to test for the
presence of a bar.   Given that the reddening is not uniform,
we will use reddening independent magnitudes that 
assume a standard extinction law for our comparison.

These magnitudes are defined as $W_V = V - 3.97\times (V - R)$. 
In the most heavily reddened fields some of the faintest variables
will be lost.  Because we only reliably detect RRab which are brighter than
$V \sim 19.5$, the distance to which we can detect
RRab with $0.4 < P < 0.7$ and $A_V > 0.2$ mag depends on the reddening.
For $E(B-V) = 0.5$ (typical of the Baade's window field), we would
detect RRab located beyond the Sgr dwarf galaxy (Paper I).

We will neglect brighter stars from our sample of bulge RR Lyrae, because they
may belong to the foreground disk. These may also be real
bulge RR Lyrae, but blended with other stars (mostly RG), which renders them
useless for our purposes. 
Likewise, we have neglected fainter RR Lyrae, because they belong to the Sgr 
dwarf galaxy, located at $D=22$ kpc behind the Galactic bulge (see Paper I).

The RR Lyrae stars are excellent distance indicators, and 
should show the effect of the bar if one is present. 
Figure 5 shows the mean reddening-independent magnitudes of all the
RRab observed. The scatter is large, but there is no clear magnitude
dependence with longitude, as one would expect from a barred
distribution. In order to quantify this, and trying to avoid the
systematic effects that may be induced when considering fields with
widely different reddenings, we study the mean magnitudes on a field by
field basis. One such systematic effect would be to miss the faintest stars in
the most reddened fields. This would raise the mean
$W_V$ of the RR Lyrae, and since the most reddened fields seem to be the inner
ones (F108, F113, F118), this could hide the presence of a bar.

We have computed the mean magnitudes of the bulge RR Lyrae stars 
in the different MACHO fields.  The lower panels of Figure 6 show 
the mean reddening-independent magnitudes of the RRab and RRc in each 
of the MACHO bulge fields. Each point comes from averaging 20-90 stars. 
RR Lyrae stars belonging to the foreground or to the Sgr
dwarf have been discarded before computing these mean magnitudes.

Table 3 lists the linear fits to different subsets of the data, of the
form $W_V = W_0 + a\times l$, including the errors in the intercept and slope,
$\sigma_W$ and $\sigma_a$, the reduced  chi-squares, and the standard
deviation  of residuals.  The best-fit line to all the stars show that
the bulk of the RR Lyrae stars do not follow the expected barred distribution.  
This is our main result, and it does not change if we eliminate the 
most reddened fields (F108, F118, F113 and F101).

Alard (1996b) also finds that the RR Lyrae population in fields at lower
latitudes monitored by the DUO experiment is not barred, and that there is
evidence for two populations.
The present result confirms  the results of
Wesselink (1987), who studied the RR Lyrae stars in low latitude fields
towards the bulge, following up the work of Oort \& Plaut (1975).
He found that there is no bar in the RR Lyrae distribution. 
\footnote{This result, quoted by 
de Zeeuw (1993), is an example of a negative result that was not published.
Negative results, though also important, are not often cited in
review articles, which only stress the positive evidence for a bar.} 
When forcing a triaxial fit to
the data, he obtained a bar slightly tilted in the opposite sense as
the bar found by Blitz \& Spergel (1991). 
That is, the RR Lyrae at positive Galactic longitudes seem to be
fainter, and therefore more distant than the ones at negative longitudes. 

\subsection{Comparison with Clump Giants}

For comparison, we have also computed the mean magnitude of the RGB clump
in each of the fields. These magnitudes are also shown in Figure 6,
and are consistent with previous results. For example, Stanek et al. 
(1994, 1996) find a difference of 0.4 mag between the mean magnitude
of clump giants in their fields at $l=-5$ and $+5$ degrees. Thus, we expect a
maximum difference in the mean magnitude of the clump of $\Delta W_V = 0.2$ 
mag between fields 124 and 157. This is observed in the MACHO data, 
as shown in Figure 6.
The slope of the solid line is that found by Stanek et al. (1996).
The difference in $W_V$ between the clump giants and RR Lyrae stars is due in
part to their color difference, because of the way this magnitude is defined.
However, from Figure 4 we can argue that there is nothing about the run of
reddening with longitude that can account for the difference between the
clump stars and the RR Lyrae stars when the mean $W_V$ are
plotted against longitude (Figure 6). 

Note that our results are independent
of the RRab absolute magnitude calibration. 
The underlying assumption is, however, that the RR Lyrae population does not
change significantly among these fields. A similar assumption is 
taken for the clump giants (Stanek et al. 1996).

\subsection{Metallicity Effects}

The abundances of RR Lyrae in the bulge range from $[Fe/H] = -1.65$
to $[Fe/H] = -0.3$, with mean $[Fe/H] = -1$ (Walker \& Terndrup 1991),
i.e. the same metallicity as the globular cluster M5 (Reid 1996). 
There is a correlation between RR Lyrae period and luminosity
with [Fe/H] (e.g. Sandage
1993, Carney et al. 1992), in the sense that the more metal--poor
RR ab tend to have longer periods.   In particular,
the morphology of the period-amplitude diagram is 
determined in part by the metallicity of the population (e.g. 
Bono \& Stellingwerf 1994).  The effect of metallicity in the
period--amplitude plane is clearly illustrated by Figs. 10--12 of
Jones et al. (1992). 

The period-amplitude diagram allows us to obtain relative abundances, and to divide the 
RRab sample into three bins, containing metal-poor, intermediate metallicity, 
and metal-rich RRab stars. Figure 7 shows how we subdivide the sample.
The most metal-rich and metal-poor RR Lyrae would be dominated by Oosterhoff (1939) types 
I, and II, respectively, found also in metal-rich and metal-poor Galactic globular clusters.
We have computed the mean $W_V$ magnitudes of these
three RRab subsamples of different metallicities.
Figure 8 shows the mean reddening-independent magnitudes of the
subsamples of different mean metallicities in each of the MACHO bulge fields. 
These do not show a barred distribution, with the
possible exception of the most metal-rich ones.
For the metal-rich subsample, the bar cannot be ruled out within the errors
listed in Table 3.

\subsection{The Inner Fields}

The inner bar in the MW extends to $l>10^{\circ}$ according to the
COBE data, and to the OGLE data. However,
the upper panel of Figure 5 shows that the bar signature in the clump
giants is stronger in the inner fields with $l<4^{\circ}$.

There is also a trend of $W_V$ with longitude in the lower panels of
Figure 5: for $l<4^{\circ}$
(i.e. $y<0.55$ kpc), the RR Lyrae stars appear to be brighter at larger longitude.
In Figure 6 we again see that for $l<4^{\circ}$, the RRc and RRab more
or less follow the trend defined by the clump stars.
The same is seen in the panels of Figure 8, though this
trend persists to large longitudes for the metal-rich RRab.

Separating the fields by latitude, in Figure 9 we plot the mean magnitudes of
the RR Lyrae types ab and c. The type c stars would be prevalent in 
Oosterhoff type I populations,
while the type ab stars would be prevalent in Oosterhoff type II populations.
We can see that the stars in the inner fields,
with $b > -4^{\circ}$ and $l< 4^{\circ}$, show a
rather compact bar, somewhat shorter than the one defined by
the clump stars but having about the same orientation.  However, the
remaining fields show no trend with longitude.

\section{Discussion}

Figure 10 shows the mean distances of RR Lyrae and clump giants on a field by field basis
projected onto the Galactic plane. The barred distribution of the clump giants, and the
absence of a bar in the bulk of the 
RR Lyrae are clear (i.e. the slope of the clump giants is significantly
negative, while the slopes of the RR Lyrae are positive).
In fact, the RR Lyrae distribution
seems to be slightly tilted in the opposite sense.  However,
within the scatter these distances are consistent with an axisymmetric distribution,
except in the inner regions as discussed in Section 4.4.

A similar effect is illustrated in Figure 11a, which shows
the $\Delta W$ $vs$ longitude, with the tangent point being the mean
of the line of sight distance distribution. Note that part of the
difference in $W_V$ is due to the color difference between RR Lyrae and
clump giants. 
We have assumed (as did Stanek et al. 1995), that there is no strong color
gradient in these populations.
Setting the zero point in order to make the mean distance of the RR Lyrae
equal to the mean distance of clump giants at $l=0^{\circ}$,
Figure 11 b shows the projected distance difference between these
populations in kpc. Even though there might be a bar structure confined
to the inner regions in the RR Lyrae, it is striking that 
the bar is stronger in
the red giant clump stars. The scatter of about $0.3$ mag
in these points correspond to about $1$ kpc.

Having established that the line of sight distribution of the bulk of
the RR Lyrae stars is not
barred, we will discuss the differences between the RR Lyrae 
and the other stellar populations that show clearly the bar.

In these inner Milky Way fields one has to consider the contribution from
all possible Galactic components, namely disk, bulge and halo. They
have different radial density profiles, and we can try to relate the RR Lyr
density distribution to one of these components. 

The RR Lyrae variables have historically been cornerstones in the understanding
of the bulge component, starting with their discovery in Baade's window by Baade (1946).
They have been used to measure the distance (Carney et al. 1995), the age 
(Lee 1992), and the metallicity of the Galactic bulge (Walker \& Terndrup 1991).  
An important problem is that the RR Lyrae stars in Baade's window
seem to be 2 Gyr older than the RR Lyrae stars in Galactic globular clusters
(Lee 1992), while the age of the dominant population of the bulge has
been measured to be as old as, or younger than globular clusters from
main-sequence turn-off photometry (Terndrup 1988, Holtzman et al. 1993, 1996,
Ortolani et al. 1995), and from the existence of long period Miras (Glass et al.
1995).  In order to solve this contradiction,
Minniti (1996) proposed that the majority of the RR Lyrae in the bulge
windows belong to the inner extension of the halo rather than to the
metal-rich bulge itself. One of the motivations for this distinction is
that the bulge RR Lyrae stars are not as metal-rich as the bulk of the bulge giants. 
(Rich 1992, Geisler \& Friel 1992, Sadler 1992, Harding \& Morrison 1993, McWilliam \&
Rich 1994).  While the mean metallicity
of K giants is $[Fe/H] = -0.25$ to $-0.6$ in bulge windows 
(Minniti et al. 1995),  the RR Lyrae stars are more 
metal-poor, with mean $[Fe/H] = -1.0$ (Walker \& Terndrup 1991).  

Here we find further evidence for the differentiation between these two
populations, namely the spatial distribution. The bulge giants, representing
the bulk of the bulge population, is barred, contrary to the RR Lyrae stars.
The metal-poor stars are not numerous in the bulge (Minniti et al. 1995).
However, the probability of the formation of RR Lyrae stars in a metal poor population
is a factor of $\sim 50$ larger than in a metal rich population (Suntzeff et
al. 1991, Layden 1995). This would explain why there are not many RR Lyrae 
that follow the clump giants in the bar.
If accurate metallicities for the present sample are measured, it may be
possible to single out the metal-rich RR Lyrae stars, and then decide if 
they are barred.
Note, for example, that the
RR Lyrae at larger distances from the Galactic plane, toward the Plaut fields,
have disk-like kinematics (Rodgers 1991). The MACHO fields, however, are
closer to the Galactic center, where the contribution from the halo and bulge
components would outnumber the disk contribution.

Since the RR Lyrae stars do not belong to the triaxial bulge, 
we argue that the majority of these RR Lyrae belong to the extension
of the Galactic halo into the inner regions.
The Milky Way halo outside of the bulge region ($R>3~ kpc$) is traced by
halo globulars (Zinn 1985), field blue horizontal branch stars 
(Preston et al. 1991), and also RR Lyrae stars (Saha 1985, Suntzeff et al. 1991).
The properties of the inner RR Lyrae are consistent with the extension of this
halo.

In particular, Gratton (1987) and Tyson (1991) measured the kinematics of 
about 30 RR Lyrae in Baade's window.   Their mean velocity dispersion 
$\sigma = 130$ km s$^{-1}$ is larger than that of the bulk of the red giants,
and consistent with a metal-poor
component (Rich 1992, Minniti 1996). 
We note that these are extremely difficult measurements made at random
phases, and that more spectroscopic observations of bulge RR Lyrae are needed.
However, the mean rotation
of this population cannot be measured in this field lying along the minor
axis, and it is still not known.  Minniti (1996) predicts that the
rotation of RR Lyrae stars would be lower than that observed for the K 
giants, as one would expect for a typical halo population, 
and Zhao et al. (1994) found that the orbits of metal-poor bulge stars
may be oriented in a sense opposite to the bar.

Should we expect to find any sign of triaxiality in a hot population like 
the RR Lyrae?  Making the approximation that the RR Lyrae are an isothermal 
population of test particles (i.e.  negligible mass), and neglecting the 
rotation of the figure of the bar, then the equidensity surfaces of the RR 
Lyrae population will follow the equipotential
surfaces of the total potential. Thus, in the absence of further kinematic
information, we can only speculate that if the potential in the inner
parts is dominated by the bar as outlined by the clump stars, then one might
expect some bar-like response to be seen in the RR Lyrae stars too. This
signature would be stronger in the inner fields, where the bar potential is
deeper, as observed. In general,
the RR Lyrae bar would be less pronounced than the clump bar for two
reasons: (1) the RR Lyrae stars are a hotter population (they have a larger 
velocity dispersion and their orbits have larger apogalactica, tending
to make their distribution more axisymmetric), and (2) the
equipotential surfaces for the underlying bar are less triaxial than
the equidensity surfaces for the bar.

To summarize, the absence of a strong bar in the RR Lyrae population is important.
The most straightforward interpretation is that they represent a different 
population than the metal-rich bulge.
The RR Lyrae stars could belong 
to the inner extension of the halo, which is relatively metal-poor, 
while the dominant metal-rich component traced by the clump giants would 
represent the bar.  

\section{Conclusions}
We have presented the mean colors and magnitudes of RR Lyrae stars in 24
bulge fields from the MACHO database.

We computed the mean reddenings for these fields, based on the mean colors. 
This allows us to identify fields with relatively low and 
uniform extinction.

The distribution along the line of sight for bulge RR Lyrae stars is examined
based on their mean magnitudes.  In particular, we searched for
evidence of a barred distribution. Taken as a whole, 
there is no bar in the RR Lyrae population
of the Galactic bulge, contrary to the evidence coming from
other tracers of metal-rich populations, such as RGB clump stars or IR sources. 
However, the bar signature is clearly seen in the mean magnitudes of clump
giants in different MACHO fields. A bar distribution can be seen 
only in the inner fields, within $l<4^{\circ}$ and $b >-4^{\circ}$, presumably
where the bar potential is strong enough to influence the kinematically
hot RR Lyrae component.

We conclude that this evidence
indicates the presence of two different populations (see also Alard 1997b).
The bulk of the
RR Lyrae represent the inner extension of an axisymmetric halo, while the
more metal-rich stars belong to a barred bulge.  However, knowledge of 
the kinematics of this RR Lyrae sample as function of metallicity 
is highly desirable,
as well as the study of fields covering a wider range of Galactic latitudes
and longitudes.

The inner RR Lyrae could be among the oldest known stars in our Galaxy,
and further studies 
would prove fruitful to our understanding of the formation of the Galaxy. 
Perhaps some of these RR Lyrae are the remains of a large putative
population of primordial globular clusters that were destroyed by 
dynamical processes in the inner regions of the Milky Way.
Note that this destructive processes would affect mostly clusters on
elongated orbits (e.g. Murali \& Weinberg 1996). If the present RR Lyrae are on
elongated orbits and do not spend enough time near perigalacticon, then
they would not feel the bar potential for long enough time to respond 
to it. Kinematic studies of the present RR Lyrae sample could help recover 
the original properties of primordial globular clusters at the time of the 
formation of the Milky Way.

\acknowledgements{
We are very grateful for the skilled support by the technical staff at Mount
Stromlo Observatory.
Work at LLNL is supported by DOE contract W7405-ENG-48.
Work at the CfPA is supported NSF AST-8809616 and AST-9120005.
Work at MSSSO is supported by the Australian Department of Industry,
Technology and Regional Development.
WJS is supported by a PPARC Advanced Fellowship.
KG thanks support from DOE OJI, Sloan, and Cottrell awards.
CWS thanks support from the Sloan, Packard and Seaver Foundations.
}

\eject

\begin{figure}
\plotone{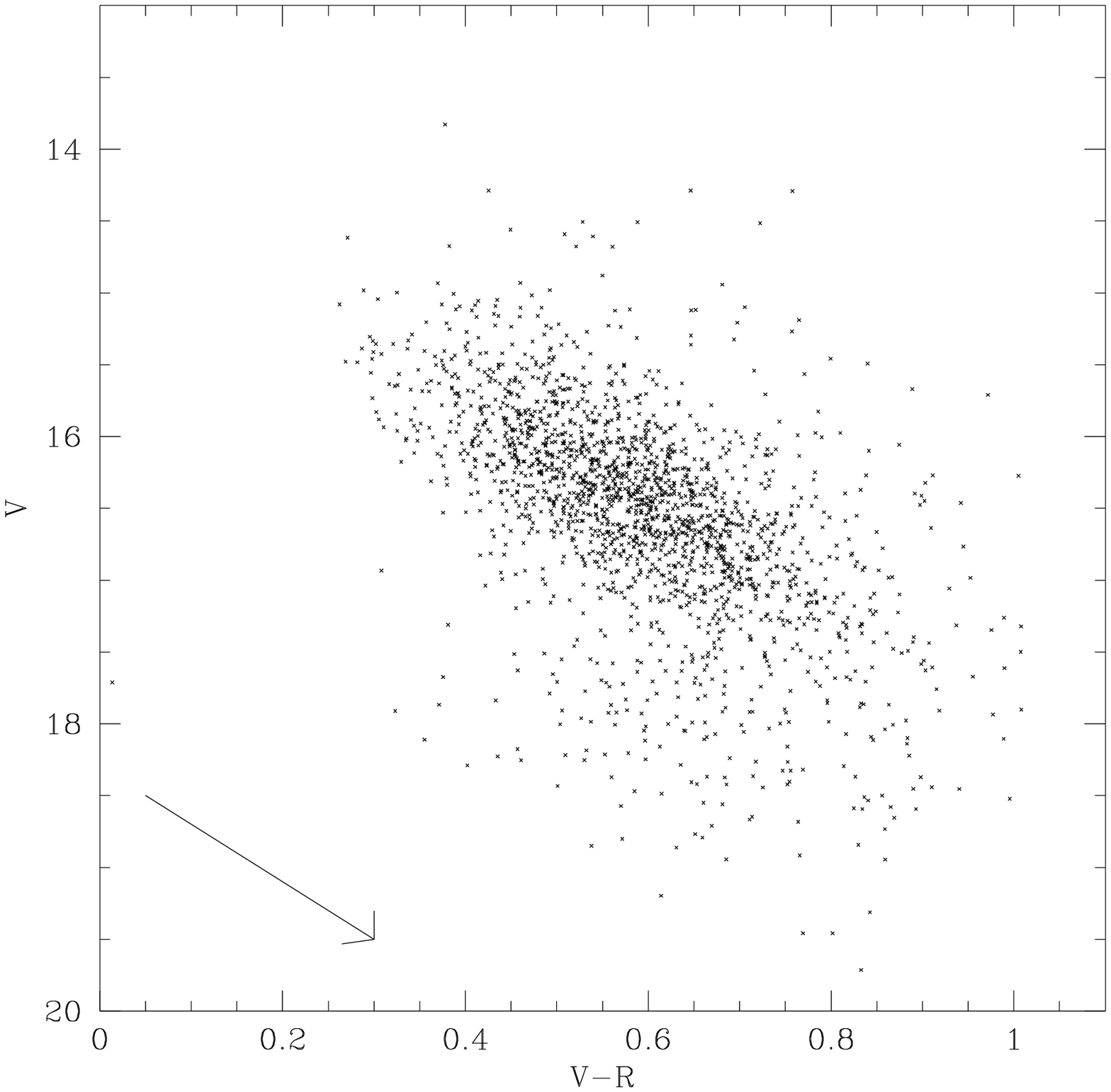}
\caption{ 
Observed V magnitude $vs.$ $V-R$
color-magnitude diagram for all bulge RR Lyr.
The fainter group of RR Lyrae belong to the Sgr dwarf galaxy (Paper I).
}
\end{figure}

\begin{figure}
\plotone{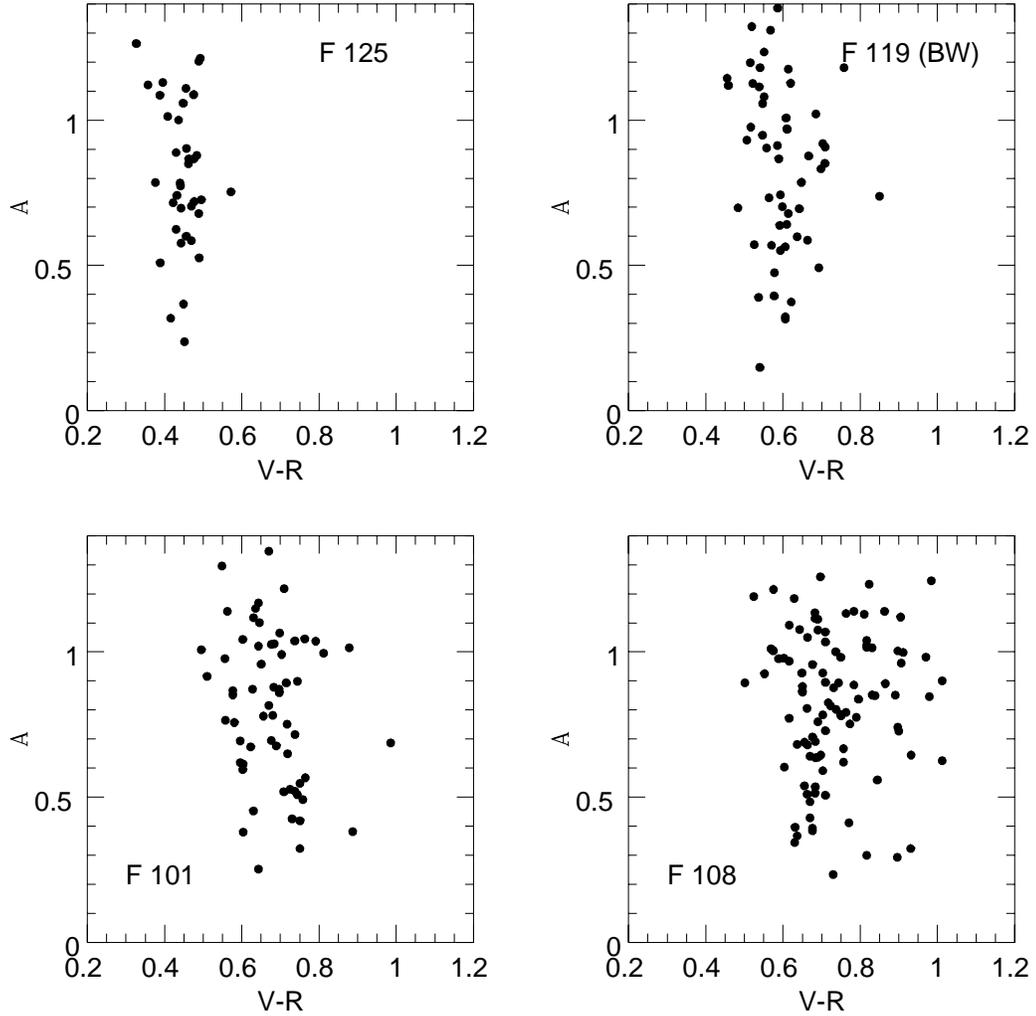}
\caption{ 
Amplitude-color diagram for four MACHO fields. 
F125 has low and uniform absorption.
F119 is Baade's window, and represents an intermediate case in terms of
the amount and homogeneity of the absorption.
F108 and F111 are very obscured, and show significant variations.
}
\end{figure}

\begin{figure}
\plotone{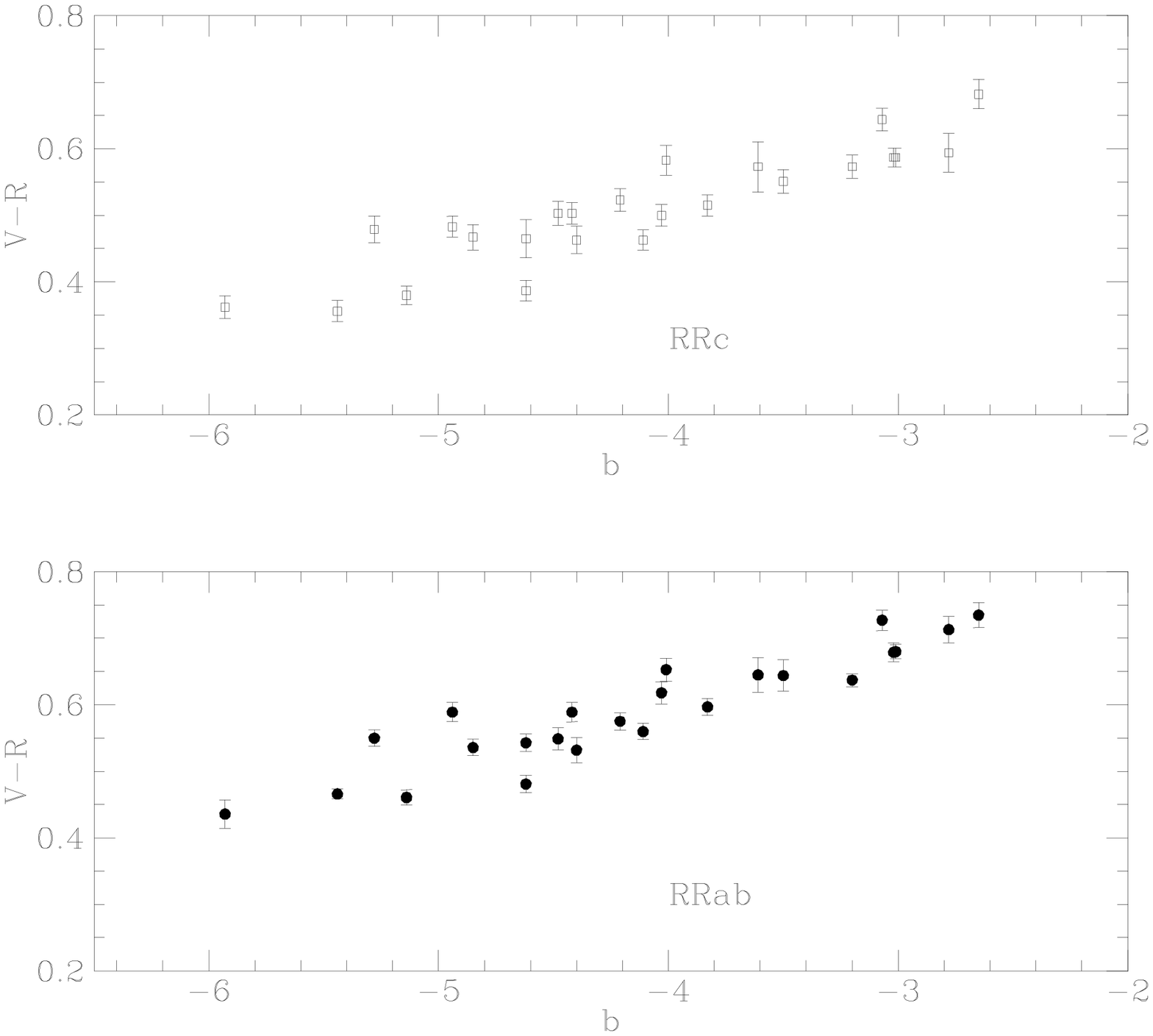}
\caption{ 
Mean colors of RRab (solid circles) and RRc (open squares)
as function of Galactic latitude.
}
\end{figure}

\begin{figure}
\plotone{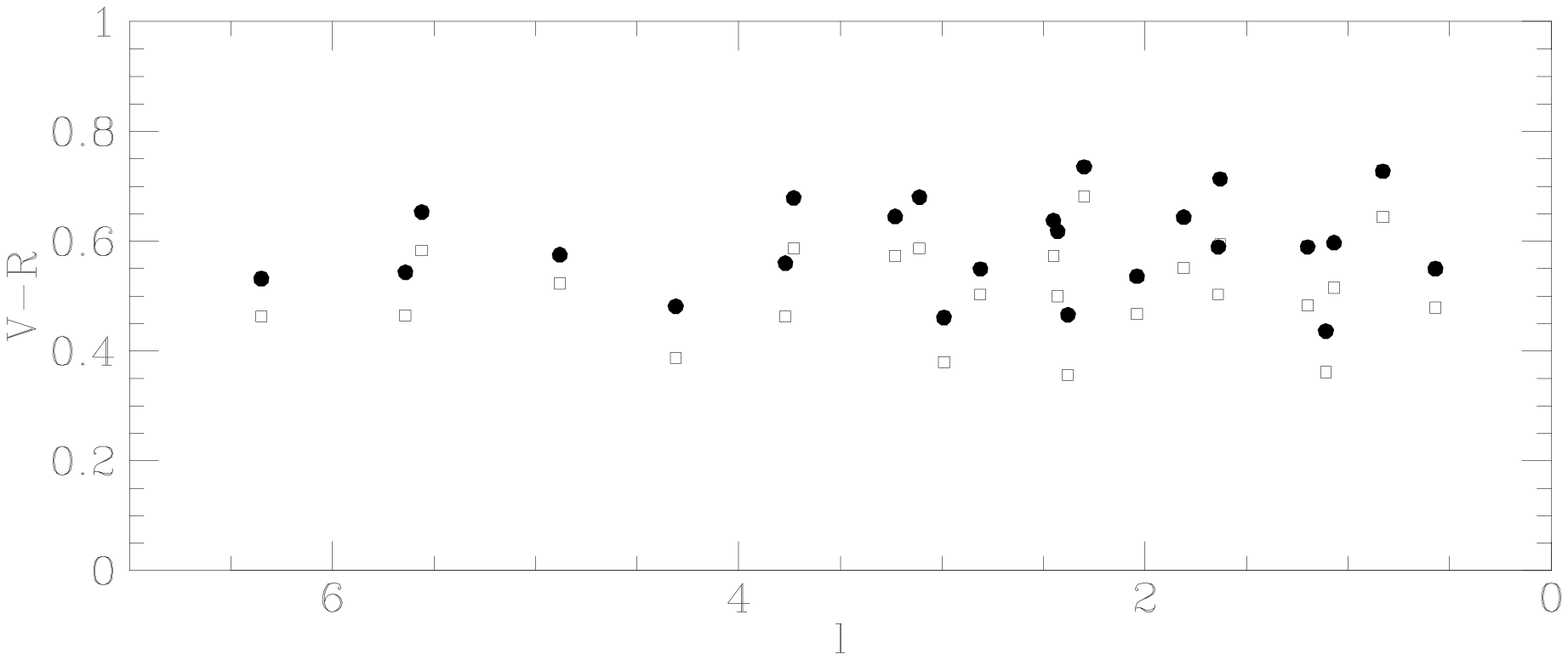}
\caption{
Mean colors of RRab (solid circles) and RRc  (open squares)
as function of Galactic longitude.
}
\end{figure}

\begin{figure}
\plotone{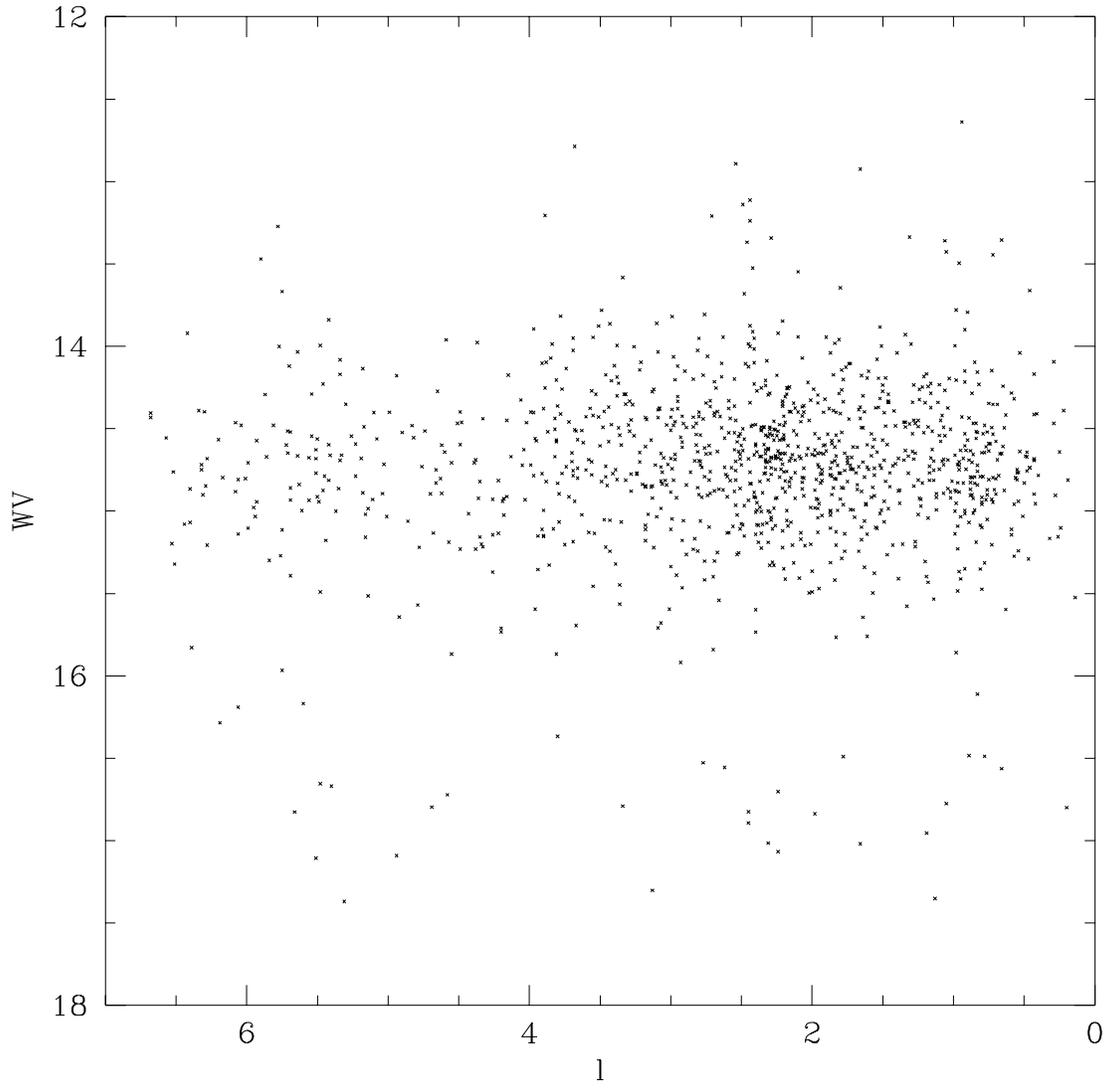}
\caption{
Mean reddening-independent magnitude $W_V$ $vs$ Galactic 
longitude for RRab stars in all the
MACHO bulge fields. There is no strong trend of magnitude with longitude.
The RR Lyrae fainter than $W_V \approx 16$ belong to the Sgr dwarf 
galaxy (Paper I).
}
\end{figure}

\begin{figure}
\plotone{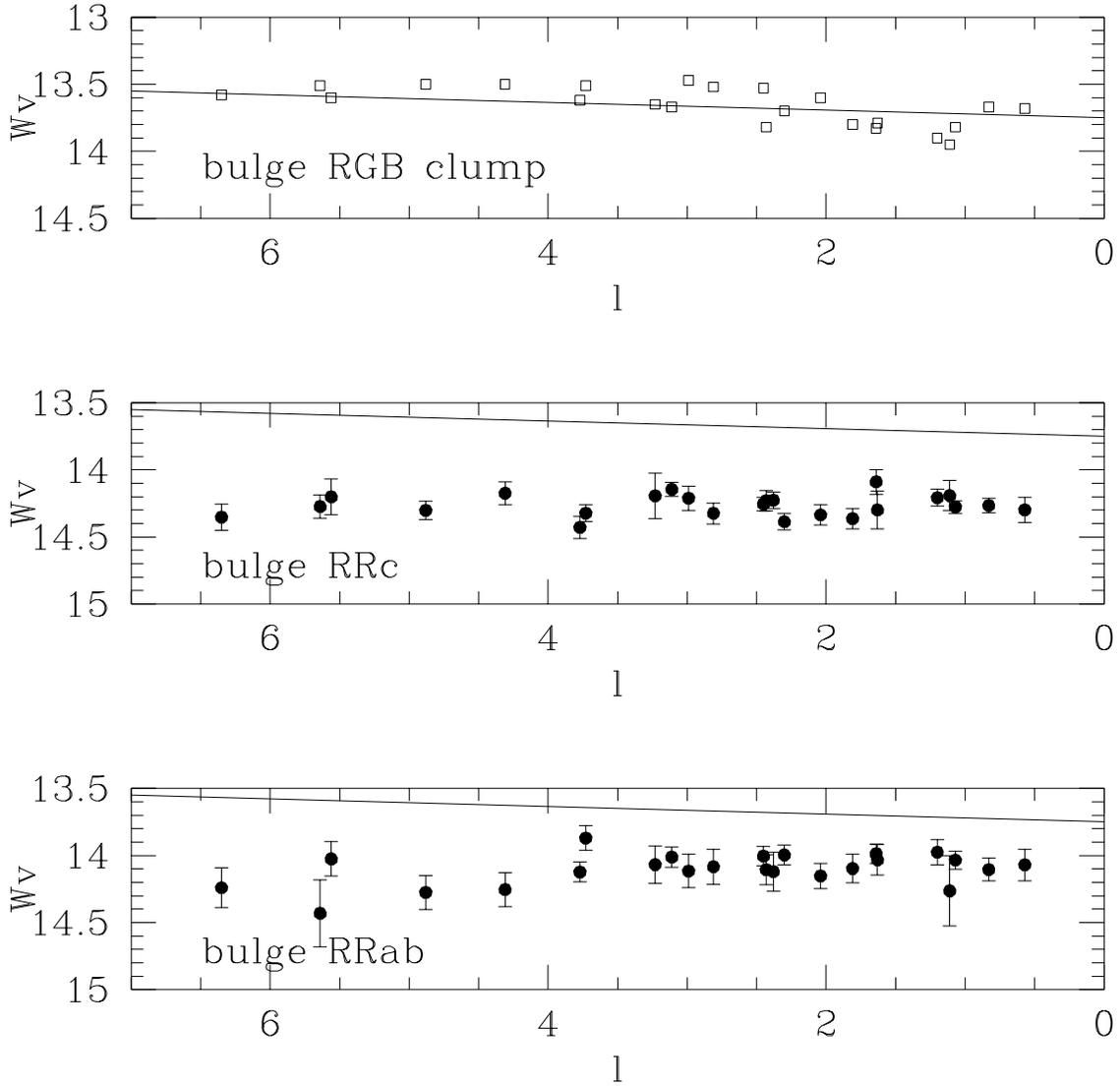}
\caption{
Mean magnitudes of RRab (bottom panel), and RRc (middle panel)
compared with the mean magnitudes of RGB clump stars (top panel) 
for the 24 MACHO bulge fields as a function of Galactic longitude.
The solid line shows the trend expected from the barred distribution
defined by the clump giants. The slope of this line is taken from
Stanek et al. (1996).  This figure shows that the RR Lyrae stars do not 
follow the barred distribution seen in the clump giants and other tracers.
}
\end{figure}

\begin{figure}
\plotone{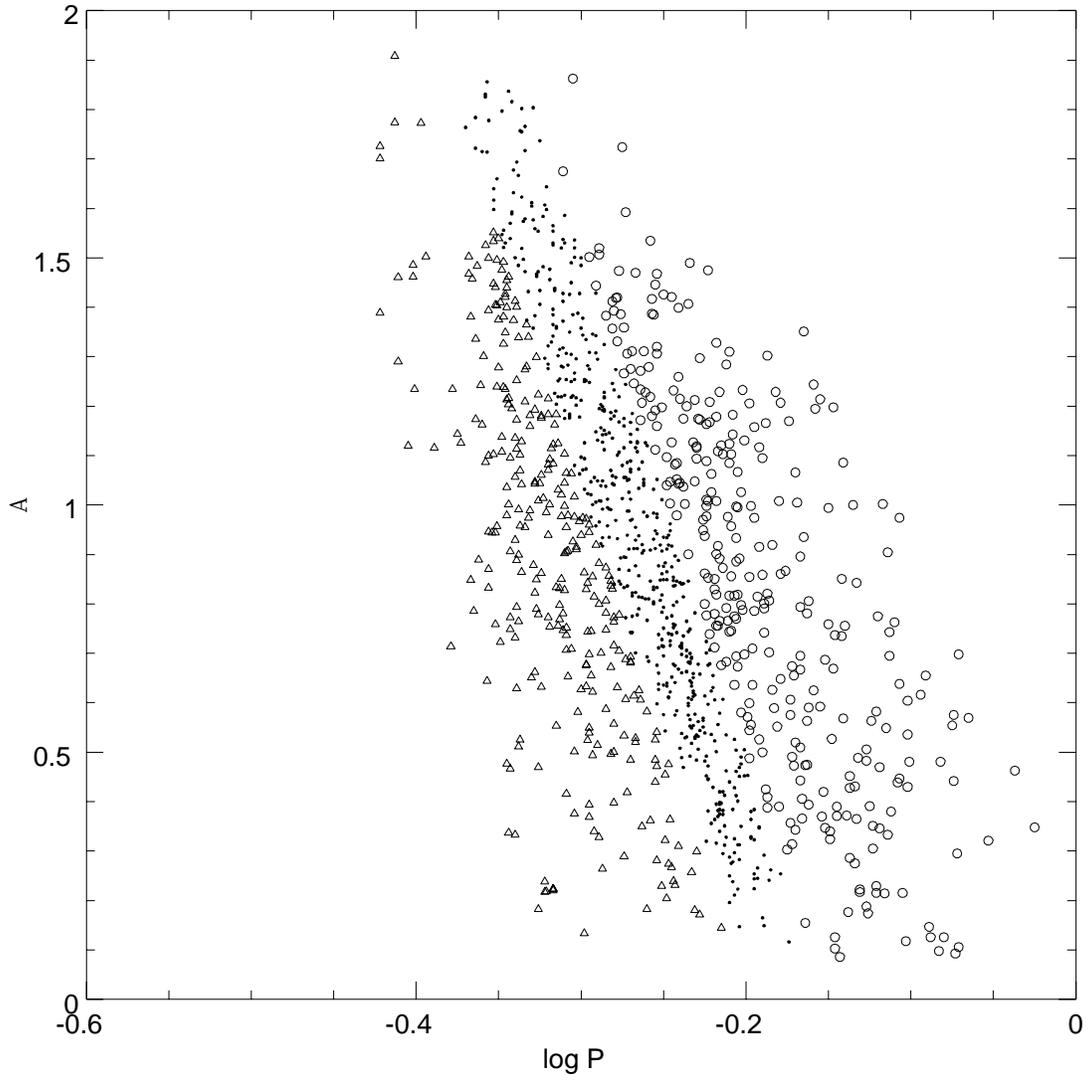}
\caption{ 
Period-amplitude diagram for RRab, showing the division into 
subsamples with metal-poor (circles),
intermediate metallicity (dots), and metal rich (triangles) stars.
}
\end{figure}

\begin{figure}
\plotone{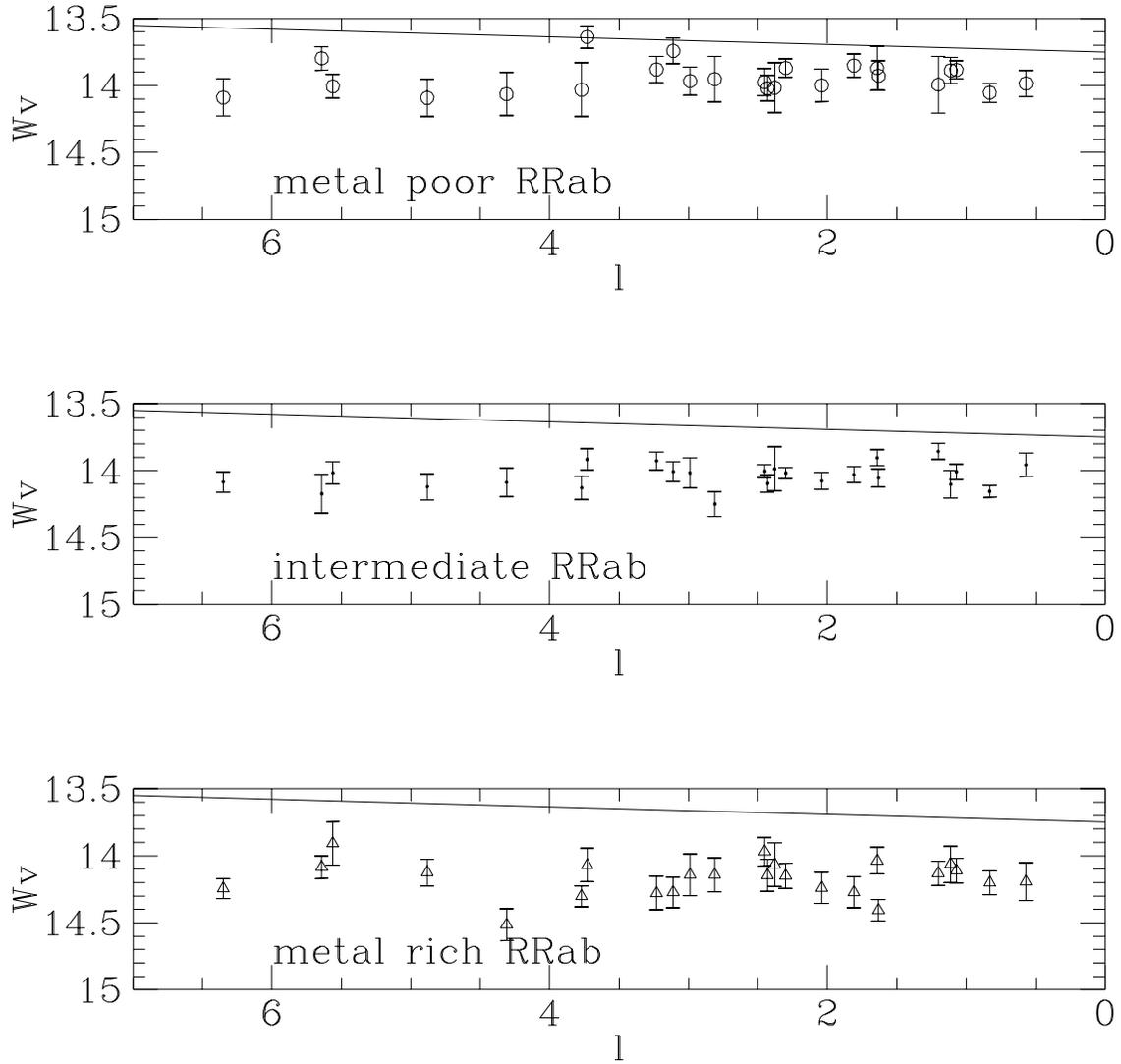}
\caption{ 
Mean magnitudes of the metal-poor (top panel), intermediate metallicity
(middle panel), and metal-rich (bottom panel) RRab $vs.$ Galactic longitude.
}
\end{figure}

\begin{figure}
\plotone{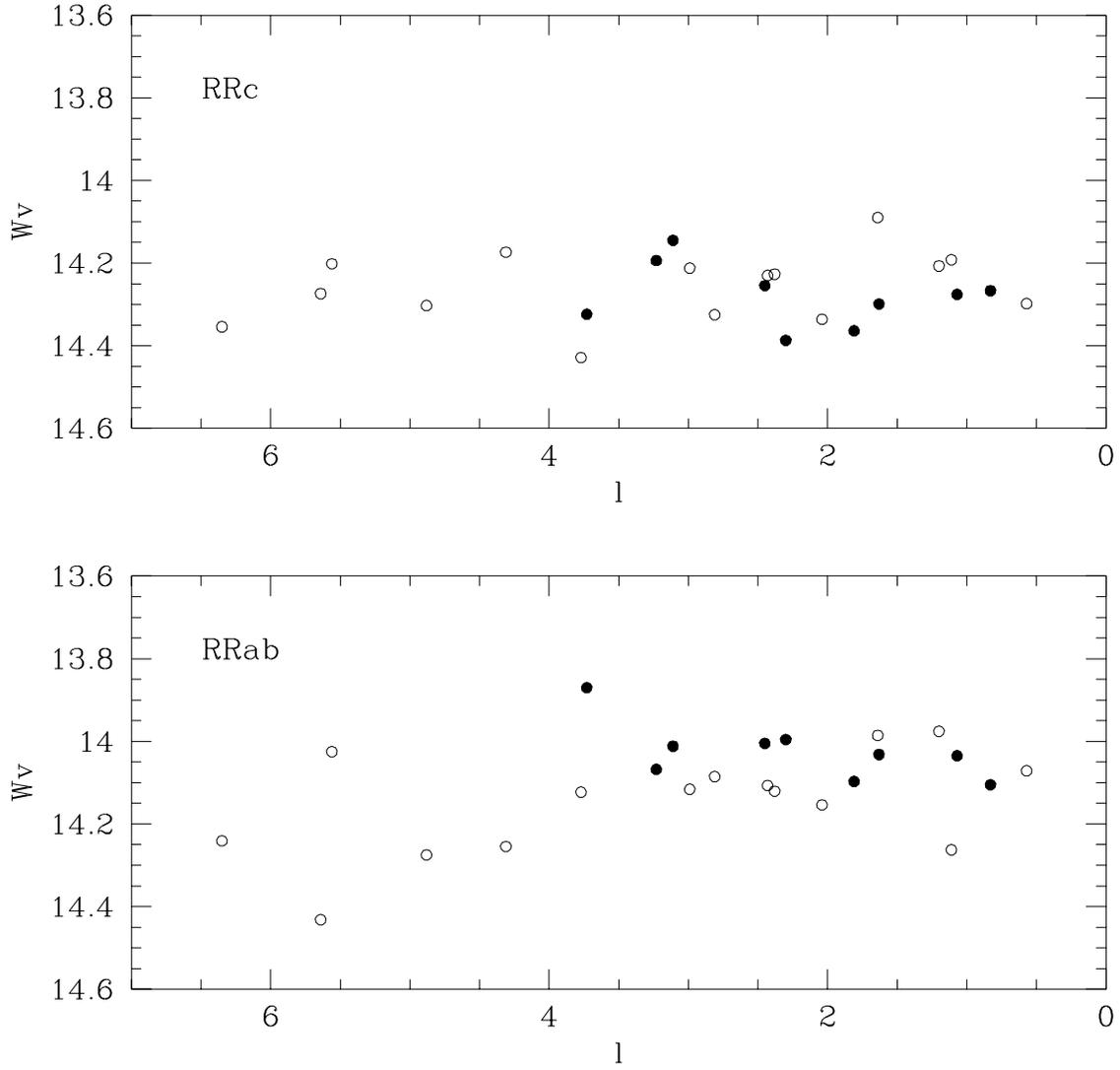}
\caption{
Mean $W_V$ magnitudes of RRLyrae stars
as function of longitude for the fields closer to
the Galactic plane with $b>-4^{\circ}$ (full circles) and lower latitude
fields (open circles).
Note the absence of a bar in the lower latitude fields, and the possible
barred distribution in the fields closer to the plane.
}
\end{figure}

\begin{figure}
\plotone{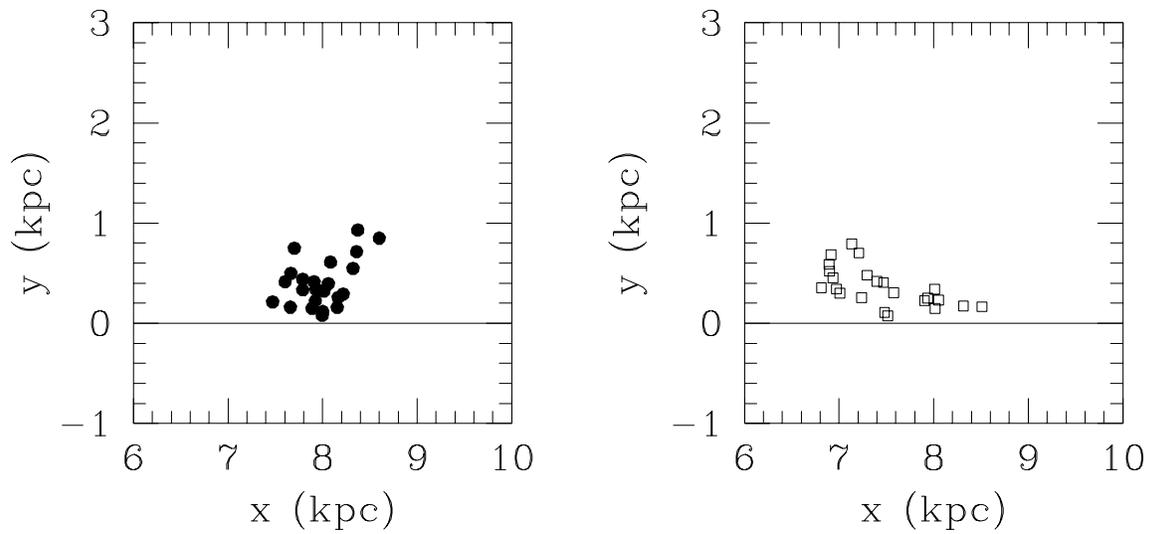}
\caption{
Mean distances in kiloparsecs plotted onto the Galactic plane.
Each point corresponds to the mean magnitude of clump giants RR Lyrae (left panel)
or clump giants (right panel) in a different bulge field.
Note the absence of a bar in the RR Lyrae, and the strong
barred distribution in the clump giants.
The solid horizontal line connects the Sun, located outside the
panels at $(0,0)$, with the Galactic center, located at $(0,8)$.
}
\end{figure}

\begin{figure}
\plotone{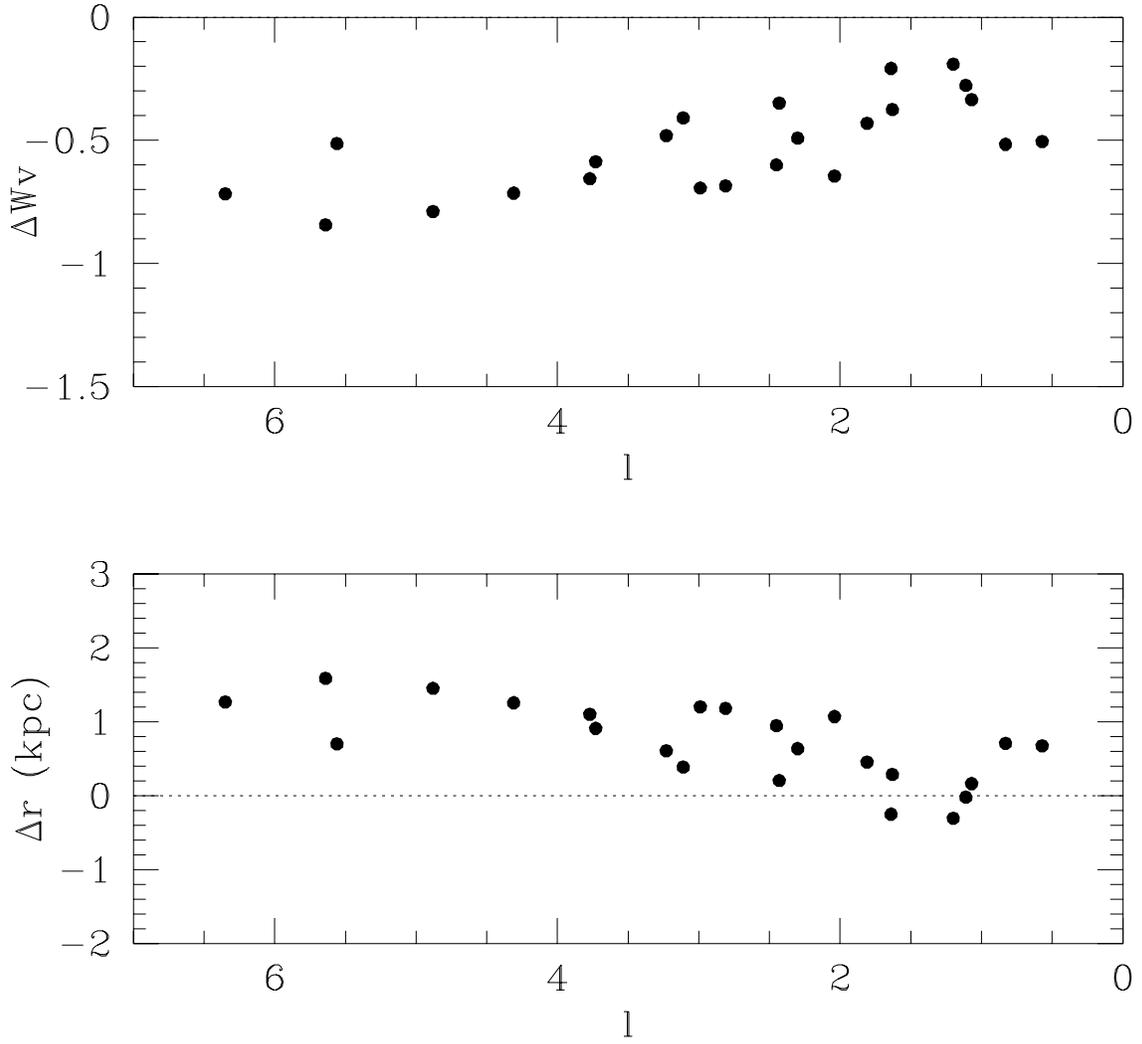}
\caption{
Difference in mean $W_V$ magnitudes (top panel), and 
distances in kiloparsecs (bottom panel) between RR Lyrae and clump giants as
function of longitude. We have assumed that $R_0 = 8$ kpc, and that
both distributions overlap at $l = 0^{\circ}$.
}
\end{figure}

\begin{deluxetable}{lrllllllll}
\small
\footnotesize
\tablewidth{0pt}
\scriptsize
\tablecaption{Location of the MACHO Bulge Fields}
\tablehead{
\multicolumn{1}{c}{Field}&
\multicolumn{1}{c}{$RA_{2000}$}&
\multicolumn{1}{c}{$DEC_{2000}$}&
\multicolumn{1}{c}{l}&
\multicolumn{1}{c}{b}&
\multicolumn{1}{c}{N*}}
\startdata
  113 & 17 57 10.1&--28 56 39 & 1.63&--2.78 & 0.296E+06\\
  118 & 17 56 30.6&--29 46 47 & 0.83&--3.07 & 0.302E+06\\
  108 & 17 58 10.9&--28 17 37 & 2.30&--2.65 & 0.255E+06\\
  109 & 18 00 40.9&--28 26 32 & 2.45&--3.20 & 0.199E+06\\
  114 & 18 00 25.7&--29 08 47 & 1.81&--3.50 & 0.208E+06\\
  104 & 18 01 23.9&--27 46 26 & 3.11&--3.01 & 0.176E+06\\
  119 & 18 00 04.4&--29 57 16 & 1.07&--3.83 & 0.204E+06\\ 
  101 & 18 02 49.5&--27 14 33 & 3.73&--3.02 & 0.146E+06\\
  105 & 18 04 02.3&--27 58 03 & 3.23&--3.61 & 0.144E+06\\
  128 & 18 03 56.4&--28 51 46 & 2.43&--4.03 & 0.151E+06\\
  120 & 18 03 44.3&--29 44 28 & 1.64&--4.42 & 0.151E+06\\
  110 & 18 06 34.0&--28 45 17 & 2.81&--4.48 & 0.123E+06\\
  102 & 18 07 11.8&--27 44 14 & 3.77&--4.11 & 0.112E+06\\
  121 & 18 04 50.4&--30 22 52 & 1.20&--4.94 & 0.132E+06\\
  115 & 18 06 22.3&--29 36 05 & 2.04&--4.85 & 0.124E+06\\
  124 & 18 04 49.2&--31 05 44 & 0.57&--5.28 & 0.122E+06\\
  167 & 18 09 58.8&--26 48 45 & 4.88&--4.21 & 0.869E+05\\
  161 & 18 10 41.1&--26 07 22 & 5.56&--4.01 & 0.778E+05\\
  103 & 18 10 22.9&--27 30 49 & 4.31&--4.62 & 0.900E+05\\
  111 & 18 09 36.0&--28 54 51 & 2.99&--5.14 & 0.100E+06\\
  116 & 18 09 29.1&--29 35 28 & 2.38&--5.44 & 0.100E+06\\
  162 & 18 13 12.7&--26 20 38 & 5.64&--4.62 & 0.699E+05\\
  125 & 18 08 41.3&--30 56 08 & 1.11&--5.93 & 0.976E+05\\
  159 & 18 13 52.6&--25 36 57 & 6.35&--4.40 & 0.630E+05\\
\enddata
\end {deluxetable}

\begin{deluxetable}{rrrrrrrrrrrrrr}
\small
\footnotesize
\tablewidth{0pt}
\scriptsize
\tablecaption{Mean magnitudes and colors for bulge RRab and RRc}
\tablehead{
\multicolumn{1}{c}{fld}& 
\multicolumn{1}{c}{Nab}&
\multicolumn{1}{c}{$V-R$}& 
\multicolumn{1}{c}{$\sigma_{V-R}$}& 
\multicolumn{1}{c}{Wv}& 
\multicolumn{1}{c}{$\sigma_W$}& 
\multicolumn{1}{c}{Nc}&
\multicolumn{1}{c}{$V-R$}& 
\multicolumn{1}{c}{$\sigma_{V-R}$}& 
\multicolumn{1}{c}{Wv}& 
\multicolumn{1}{c}{$\sigma_W$}& 
\multicolumn{1}{c}{$E_{V-R}$}}
\startdata
101&  61&0.679&0.089&  13.870&0.566&  46&0.587& 0.090&  14.362&0.432&0.48\\
102&  33&0.560&0.055&  14.123&0.338&  23&0.463& 0.068&  14.479&0.387&0.35\\
103&  36&0.481&0.056&  14.255&0.535&  18&0.387& 0.066&  14.174&0.361&0.27\\
104&  52&0.680&0.069&  14.012&0.490&  45&0.587& 0.092&  14.185&0.375&0.47\\
105&  54&0.645&0.096&  14.068&0.517&  15&0.573& 0.142&  14.221&0.622&0.43\\
108& 103&0.735&0.111&  13.996&0.429&  39&0.682& 0.131&  14.409&0.378&0.51\\
109&  63&0.637&0.063&  14.005&0.447&  42&0.573& 0.106&  14.260&0.367&0.44\\
110&  46&0.549&0.082&  14.085&0.635&  25&0.503& 0.086&  14.349&0.392&0.33\\
111&  28&0.461&0.054&  14.116&0.603&  25&0.380& 0.067&  14.231&0.426&0.26\\
113&  57&0.713&0.077&  14.032&0.437&  16&0.594& 0.112&  14.326&0.535&0.49\\
114&  69&0.644&0.116&  14.097&0.522&  26&0.551& 0.089&  14.388&0.374&0.40\\
115&  44&0.536&0.063&  14.154&0.480&  28&0.467& 0.097&  14.256&0.478&0.34\\
116&  28&0.466&0.038&  14.121&0.745&  26&0.356& 0.080&  14.227&0.320&0.26\\
118&  82&0.727&0.091&  14.105&0.502&  40&0.644& 0.102&  14.310&0.348&0.50\\
119&  62&0.597&0.078&  14.035&0.419&  43&0.515& 0.100&  14.241&0.319&0.40\\
120&  41&0.589&0.074&  13.986&0.370&  26&0.503& 0.084&  14.090&0.471&0.38\\
121&  42&0.589&0.065&  13.976&0.430&  24&0.483& 0.072&  14.263&0.313&0.37\\
124&  40&0.550&0.056&  14.071&0.522&  21&0.479& 0.091&  14.304&0.411&0.35\\
125&  38&0.436&0.071&  14.263&0.870&  12&0.362& 0.056&  14.189&0.354&0.25\\
128&  57&0.618&0.087&  14.107&0.580&  31&0.500& 0.084&  14.286&0.424&0.40\\
159&  34&0.532&0.063&  14.241&0.491&  12&0.463& 0.068&  14.384&0.326&0.34\\
161&  39&0.653&0.070&  14.025&0.516&  16&0.583& 0.090&  14.202&0.536&0.44\\
162&  26&0.543&0.054&  14.432&1.031&  17&0.465& 0.118&  14.274&0.354&0.34\\
167&  38&0.575&0.067&  14.275&0.631&  26&0.523& 0.084&  14.290&0.338&0.38\\
\enddata
\end {deluxetable}

\begin{deluxetable}{llrlllllll}
\small
\footnotesize
\tablewidth{0pt}
\scriptsize
\tablecaption{Fits to RR Lyrae Distribution}
\tablehead{
\multicolumn{1}{c}{Stars}&
\multicolumn{1}{c}{$W_0$}&
\multicolumn{1}{c}{$a$}&
\multicolumn{1}{c}{$\sigma_W$}&
\multicolumn{1}{c}{$\sigma_a$}&
\multicolumn{1}{c}{$\chi$}&
\multicolumn{1}{c}{$res$}}
\startdata
All unbinned      &14.696&  0.000&0.027 &0.011&0.147&0.384\\
All $b<-4$        &14.654  &0.021&0.034 &0.010&0.151&0.388\\
All $b>-4$        &14.750&--0.025&0.036 &0.014&0.147&0.383\\
All   $b>-4$, $l<4$&14.821&--0.061&0.040&0.017&0.143&0.378\\
RGB  clump        &13.824&--0.057&0.021 &0.007&4.823&0.107\\
Binned RRab       &14.008&  0.022&0.030 &0.011&1.546&0.116\\
Binned RRc        &14.253&  0.006&0.031 &0.011&1.178&0.080\\
RRab metal poor   &13.937&--0.008&0.040 &0.013&1.357&0.117\\
RRab metal int.   &14.015&  0.007&0.028 &0.010&1.711&0.090\\
RRab metal rich   &14.179&  0.003&0.042 &0.012&1.581&0.135\\
\enddata
\end {deluxetable}

\end{document}